# The effects of packing structure on the effective thermal conductivity of granular media: A grain scale investigation

Weijing Dai[1], Dorian Hanaor[2], Yixiang Gan[1,*]

[1] School of Civil Engineering, The University of Sydney, NSW 2008, Australia.

[2] Fachgebiet Keramische Werkstoffe, Technische Universität Berlin, Germany.

* Corresponding author: yixiang.gan@sydney.edu.au.

**Abstract**: Structural characteristics are considered to be the dominant factors in determining the effective properties of granular media, particularly in the scope of transport phenomena. Towards improved heat management, thermal transport in granular media requires an improved fundamental understanding. In this study, the effects of packing structure on heat transfer in granular media are evaluated at macro- and grain-scales. At the grain-scale, a gas-solid coupling heat transfer model is adapted into a discrete-element-method to simulate this transport phenomenon. The numerical framework is validated by experimental data obtained using a plane source technique, and the Smoluschowski effect of the gas phase is found to be captured by this extension. By considering packings of spherical $SiO_2$ grains with an interstitial helium phase, vibration induced ordering in granular media is studied, using the simulation methods developed here, to investigate how disorder-to-order transitions of packing structure enhance effective thermal conductivity. Grain-scale thermal transport is shown to be influenced by the local neighbourhood configuration of individual grains. The formation of an ordered packing structure enhances both global and local thermal transport. This study provides a structure approach to explain transport phenomena, which can be applied in properties modification for granular media.

**Keywords**: granular media, packing structure, grain-scale, effective thermal conductivity, discrete element method.

**Highlights**

1. A numerical framework combining finite element analysis and discrete element simulation is established to study heat conduction in granular media.

2. Finite element analysis is used to correct the conventional analytical solution and an empirical correlation is provided to implement the correction.

3. The modified heat transfer model is incorporated into discrete element simulation.

4. The effects of structural transitions on effective thermal conductivity is evaluated by grain-scale structure index, and meso-scale mechanisms of the effect are identified.

**Nomenclature**

Geometrical parameters of contact units

| | | | |
|---|---|---|---|
| $r_i, r_j$ | Radius of grains | | |
| $r_{ij}$ | Effective radius of grain-grain contacts | | |
| $r_{min}$ | Minimum radius | | |
| $h$ | Separation distance | | |
| $D_{ij}$ | Distance between two centroids | | |
| $d$ | Indentation depth | | |
| $r_{cont}$ | Contact radius | | |
| $r_{iw}$ | effective radius of grain-wall contacts | | |
| $r_{eff}$ | Effective radius of cylindrical thermal resistors | | |
| $\chi$ | Ratio between $r_{eff}$ and $r_{min}$ | | |
| $L_f$ | Feature length of interstitial space | | |

Parameters used in finite element simulation

| | |
|---|---|
| $r_c$ | Radius of grains |
| $\Delta T$ | Temperature difference |
| $H_{FEM}$ | Heat flow |
| $H_{BOB}$ | Heat flow calculated by Batchelor & O'Brien model |

Parameters used in discrete element simulation

| | |
|---|---|
| $H_{inflow}$ | Inflow heat of entire media |
| $L_{axis}$ | Axial height |
| $\Delta T_{grap}$ | Apparent temperature gap |
| $A_{cross-section}$ | Cross-sectional area |
| $H_i$ | Heat flow of individual contacts |
| $H_{flow}$ | Total heat flow of grains |
| $\Delta T_{gradient}$ | Fitted temperature gradient |

Parameters used to determined gas thermal conductivity

| | | | |
|---|---|---|---|
| $k_g$ | Thermal conductivity | $k'_g$ | $k_g$ by the Smoluschowski effect |
| $k_B$ | Boltzmann constant | $P_g$ | Gas pressure |
| Kn | Knudsen number | $d_g$ | Kinetic molecular diameter |
| $\gamma$ | Energy transfer coefficient of gas | $T$ | Thermodynamic temperature |
| $m_s$ | Molar mass of solid | $m_g$ | Molar mass of gas |
| $m_r$ | Molar mass ratio $m_s/m_g$ | | |

Parameters used to characterise heat conduction and packing structure of granular media

| | | | |
|---|---|---|---|
| $k_{eff}$ | Effective thermal conductivity | $\varphi$ | Packing fraction of granular media |
| $k_s$ | Thermal conductivity of solid grain | $V_{Voronoi}$ | Volume of Voronoi cells |
| $\alpha$ | Thermal conductivity ratio $k_s/k_g$ | $S_6$ | Structural index |
| $H_{cont}$ | Contact heat conductance | $\boldsymbol{x}_i$ | Contact vector |
| $R_{cont}$ | Contact thermal resistance | $\boldsymbol{k}^*$ | Thermal conductivity tensor of grains |
| $R_{total}$ | Total thermal resistance of units | | |
| $R_{cyl.i}$, $R_{cyl.j}$ | Thermal resistance of cylindrical thermal resistors | $\Delta \boldsymbol{T}_{gradient}$ | Temperature gradient vector |
| | | $k_{zz}$ | $zz$ component in $\boldsymbol{k}^*$ |
| $\boldsymbol{H}_{density}$ | heat flux density vector | $\theta$ | Thermal conductivity anisotropy |

# 1 Introduction

Effective thermal management and waste heat dissipation are keys to improving efficiency in diverse industrial applications [1-3]. Towards this end, granular media are extensively used and are often studied by researchers in disciplines of thermal science and engineering [4-7]. Understanding and managing heat transfer processes in granular media necessitates the accurate evaluation of their effective thermal conductivity. In a broader context, granular media are regarded as multi-phase substances of which the effective thermal conductivity can be described by the Hashin-Shtrikman bound, the Maxwell's equation and the effective medium theory [8]. However, these methods do not provide sufficient mechanistic insights, motivating further fundamental studies into thermal transport through granular media.

For general binary solid-gas or solid-fluid granular media, major heat transfer pathways include: (1) heat conduction through solid particles, (2) heat transfer through solid-solid interfaces, (3) heat transfer through gas/fluid films near particle surfaces, (4) heat radiation between solid surfaces in solid-gas cases and from solid surfaces into nearby liquids in solid-fluid cases in high temperature condition ($T > 450$K), (5) convective heat transfer between solid and mobile gas/fluid, (6) natural convection in solids and fluids of high Rayleigh number, (7) heat transfer by macroscopic fluid flow [9], and (8) granular convective heat transfer due to the motion of solid [5]. Analytically, representative geometry methods have been widely used [10, 11], which combine contributions from these aforementioned heat pathways according to the corresponding artificial geometries. In bi-phasic systems, cylinders [10, 11] and cubic [12] have been used to determine the relative contributions of each pathway. To account for the dissimilarity between the representative geometries and the meso-scale structure of granular media, including grain size distribution, contact areas, and porosity, adjustable parameters are frequently applied. But such parameters are generally too coarse or artificial to meaningfully describe granular packing structures. Rather than using representative geometries, unit cell methods that derive effective thermal conductivity by integrating grain-scale contact unit cells [13-17] which are commonly formed two adjacent grains have also been developed. One advantage of unit cell methods is that they facilitate the incorporation of grain-scale heat transfer mechanisms along with appropriate abstractions of packing structure, such as coordination number [13, 17]. But similar to the representative geometry methods, it is difficult to define irregular packing structures using quantifiable parameters [18]. With sufficient computational power, discrete element methods become favourable to study realistic irregular packing structures [19-21] for heat transfer purpose. One solution is to generate realistic packing structures to justify and improve the accuracy of model parameters used in the analytical methods [22]. The other way is to directly simulate the heat transfer process by implementing the apt grain-scale thermal interactions [23-26], such as the models developed in the unit cell methods [13, 27-30].

Although the well-established heat transfer theories of continuum media have been successfully adapted to describe granular media without missing the characteristics of packing structure [31-33] by the discrete element methods, how to explain the effects of these characteristics is still unclear. By employing the grain-scale packing fraction as one of the variables [27, 34, 35] to derive heat transfer models, the importance of grain-scale packing structure is proved. Nonetheless, effects of other structural characteristics remain inconclusive. It has been shown that the packing structure is of equivalent importance as packing fraction [36], but the question regarding which structural indicators are appropriate for the quantitative characterisation of granular packings in heat transfer remains unanswered. Attempts has been made to investigate the influence of grain-scale stress network on thermal transport [37], but the link between such networks and granular packing structure requires further study [38]. Relationships have been examined between effective mechanical properties and the grain-scale order level of the packing structures [39, 40]. Whether similar relationships can be established in the heat transfer context has not yet been discussed. The disorder-to-order transition of granular media, in which regular packing structures are formed, has been recognised as one of the major events in agitated states [41]. It is therefore worth examining in depth the relationship between the heat transfer and the order of packing structure, which can further reflect the effects of external agitation on the effective thermal conductivity.

In the current study, the open source discrete element software LIGGGHTS [42] is applied in a modified approach to investigate packing and heat transfer in a generalised static solid-gas granular media. It is demonstrated that disorder-to-order transitions induced by vibration enhance the effective thermal conductivity of granular media as ordered packing structures lead to higher grain-scale thermal conductivity.

## 2 Grain-scale heat transfer model

In this work, a modified Batchelor & O'Brien model is employed to describe inter-granular heat transfer. Contact units comprising two contacting or separated hemispheres form the basis of the grain-scale model. Here, with finite element analysis, a power function is numerically derived to extend the original method, including the correction of overestimation and the addition of the Smoluschowski effect.

### 2.1 Basic solution

The original Batchelor-O'Brien (BOB) model only considers granular media with large ratio $\alpha$ of solid thermal conductivity $k_s$ to gas thermal conductivity $k_g$, i.e., $k_s/k_g \gg 1$ [13]. So the temperature inside individual grains can be assumed as constant.

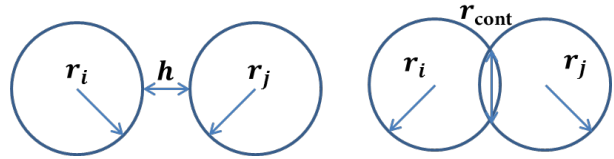

Figure 1 Separated (left) and contacting (right) pairs of grains.

The heat transfer equations determined by this condition have been numerically solved according to the separation distance $h$ or the radius of contact surface $r_{\text{cont}}$, as depicted in Figure 1. When two grains are separated with $h \geq 0$, heat conductance $H_{\text{cont}}$ of one unit is

$$H_{\text{cont}} = \begin{cases} \pi\, k_g\, r_{ij} \ln \alpha^2 & \lambda < 0.01 \\ \pi\, k_g\, r_{ij} \ln\left[1 + \dfrac{r^{*2}}{r_{ij} h}\right] & \lambda > 100 \\ \text{Minimum between the above two} & \text{else} \end{cases} \qquad 2\text{-}1$$

In the above formulas, $\alpha = k_s/k_g$, $r_{ij} = 2\, r_i\, r_j/(r_i + r_j)$ and $\lambda = \alpha^2\, h/(r_{ij}\, r^*)$ have no exact values but must comply with $r^* \gg \sqrt{r_{ij}\, h}$ [13]. So $r^*$ is chosen to be $r_{ij}$ in this work. Furthermore, a cutoff value $h_{\text{cutoff}}$ is required to avoid considering exceeded pairs of separated. It has been shown that $\varepsilon = 0.5$ in $h_{\text{cutoff}} = \varepsilon\, r_{ij}$ is suitable for general consideration [43]. If two grains are in contact at a circular contact patch having $r_{\text{cont}}$, $H_{\text{cont}}$ of this unit becomes

$$H_{\text{cont}} = \begin{cases} \pi\, k_g r_{ij} \left[\frac{2\beta}{\pi} - 2\ln\beta + \ln\alpha^2\right] & \text{if } \beta > 100 \\ \pi\, k_g r_{ij} [0.22\, \beta^2 - 0.05 \ln\beta^2 + \ln\alpha^2] & \text{if } \beta < 1 \\ \text{linear interpolation} & \text{else} \end{cases} \quad 2\text{-}2$$

where $\beta = \alpha\, r_{\text{cont}}/r_{ij}$ indicates the contribution of contact areas. In combination with Hertzian contact mechanics, $r_{\text{cont}}$ of contact surface is calculated by the distance $D_{ij}$ between centroids of two grains $i$ and $j$ as well as the indentation depth $d = (r_i + r_j) - D_{ij}$ according to $r_{\text{cont}} = \sqrt{r_{ij}\, d/2}$. Reciprocally, every contact unit can be regarded as a thermal resistor $R_{\text{cont}} = 1/H_{\text{cont}}$. However, in realistic scenarios the condition of $k_s/k_g \gg 1$ is rarely satisfied, e.g., glass vs. air, so thermal contact resistors tend to be underestimated. To make the model applicable in more practical scenarios, the temperature variation inside individual grains cannot be neglected. In the modified BOB model, two cylindrical thermal resistors $R_{\text{cyl.}\, i\,\&\, j}$ of identical $k_s$ are serially connected to the thermal contact resistor [28], compensating for the non-inclusion of the thermal resistance of solid hemispheres in each contact unit. Finally, the total thermal resistance of one contact unit is formulated as $R_{\text{total}} = R_{\text{cont}} + R_{\text{cyl.}\, i} + R_{\text{cyl.}\, j}$. In this way, the thermal resistance of cylinders is calculated according to their geometry as

$$R_{\text{cyl.}\, i\, \text{or}\, j} = \frac{r_{i\, \text{or}\, j}}{\pi\, k_s\, r_{\text{eff}}^2}. \quad 2\text{-}3$$

where the $r_{\text{eff}}$ is the radius of the cylinders and is further defined as $r_{\text{eff}} = \chi \min(r_i, r_j)$ for a more general poly-dispersed scenario. Therefore, the determination of $\chi$ becomes the essential step towards ensuring the accuracy of the modified (BOB) model. In reported methods, $\chi$ is empirically fitted according to experimental data [26, 28, 44]. Here, a finite element method is introduced to numerically calculate $\chi$. In addition, a flat wall can be approximated as a sphere with infinitely large radius for the grains pairing with flat wall boundaries, so $r_{ij}$ is replaced by $r_{iw} = \lim_{r_w \to \infty} (2\, r_i\, r_w/r_i + r_w) = 2r_i$. The other principles are still applied.

## 2.2 Determination of grain-scale parameters

As discussed in the previous section, $\chi$ is the only parameter to be determined for intergranular contacts. In the literature, $\chi = 0.5$ is used for $\alpha = 120$ [28] and $\chi = 0.71$ is used for $\alpha = 10$ [26], both giving good agreement with experimental results. But it is prohibitively time consuming to collect experimental data across the entire range of $\alpha$ values, not to mention the associated measurement uncertainty. Therefore, an estimation of $\chi$ is useful for prediction purposes. Here, finite element analysis (using ABAQUS software) is employed [17, 45] to evaluate $\chi$. In finite element simulations, contact units are assumed to be cylindrical, so axisymmetric modelling is used to reduce computational cost.

Two representative quadrants (solid) of $k_s = 100$ W/mK with the interstitial part (gas) of $k_g = k_s/\alpha$ form separated and contacting contact units as shown in the left and right of Figure 2, respectively. The radii of the quadrants are set to be $r_c = 1$ m and the separation distance $h$ and the contact radius $r_{cont}$ are varied to generate different geometries for simulations. Since material properties have no size-dependence, the exact dimensions can be later normalized. In the contacting scenario, the contacting surface is created by truncating these quadrants. The difference of the remaining height between the Hertzian quadrants and the current one is negligible (less than 0.1%) if the contact units are not severely compressed (indentation smaller than 0.2%). A constant temperature difference $\Delta T$ of 1K is imposed between the top and bottom surfaces of each contact unit while the axial side boundaries are set to adiabatic. The interfaces between solid and gas as well as between solid and solid are tied so no temperature difference exists. $\alpha$ is varied from $10^6$ to 4 because the original BOB model gives a contradictory trend when $\alpha$ is larger than 4. $r_{cont}/r_c$ is varied from 0.0477 to 0.002 and $h/r_c$ is varied from 0.000002 to 0.2. Examples of the simulated temperature fields are shown in Figure 2. The total heat flow in the simulation $H_{FEM}$ passing through the top surface is extracted at steady-state and compared with the heat flux calculated based on the original BOB model $H_{BOB}$. In this way the $\chi$ is evaluated by

$$\chi = \sqrt{\left(\frac{\Delta T}{H_{FEM}} - \frac{\Delta T}{H_{BOB}}\right) \frac{\pi k_s r_i}{2}}^{-1} . \qquad 2\text{-}4$$

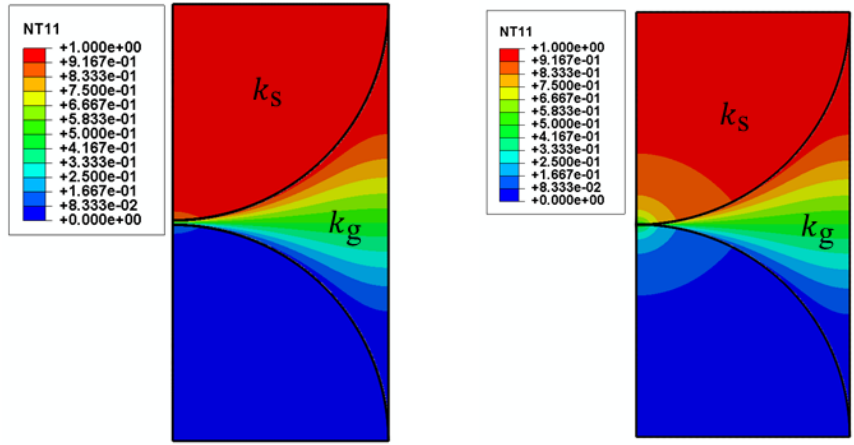

Figure 2 Separated (left) and contacting (right) geometries coloured with steady state temperature field in the finite element simulation.

In this work, the finite element analysis is considered to be more accurate than the original BOB model due to its capability to simulate temperature variation within grains. Both heat flows, $H_{\text{BOB}}$ and $H_{\text{FEM}}$, are further nondimensionlised by a factor of $(\Delta T k_s r_i)$ in the following discussion. According to (a) and (b) in Figure 3, the original BOB model shows fairly good accuracy for $\alpha \geq 10000$, but the overestimation turns out to be large when $\alpha \leq 1000$. Therefore, the evaluation of $\chi$ is mainly conducted in the $\alpha$ range from 1000 to 4 which is common practice, as shown in (c) and (d) in Figure 3. For the contacting units, $\chi$ becomes stable against $r_{\text{cont}}$ as $\alpha \leq 200$, implying the compression on the contact units has limited influence on $\chi$ in this condition. Although the model underestimates the heat flow and gives negative $\chi$ in the separated scenario as $\alpha \geq 200$, it occurs at relatively large $h/r_c$ and results in small heat flow. Thereby, this negative part is neglected, and $\chi$ of corresponding $\alpha$ in both scenarios is comparable.

The representative $\chi$ for each $\alpha$ is calculated as the mean over all $r_{\text{cont}}/r_c$ and $h/r_c$ cases for contacting and separated spheres, respectively. It can be seen from (e) in Figure 3 that $\chi$ exhibits similar values in both scenarios. Thus, the average $\chi$ of two scenarios is considered, and finally a fitting power function is obtained to estimate $\chi$ against $\alpha$ as,

$$\chi = 1.3121\, \alpha^{-0.19}. \qquad\qquad 2\text{-}5$$

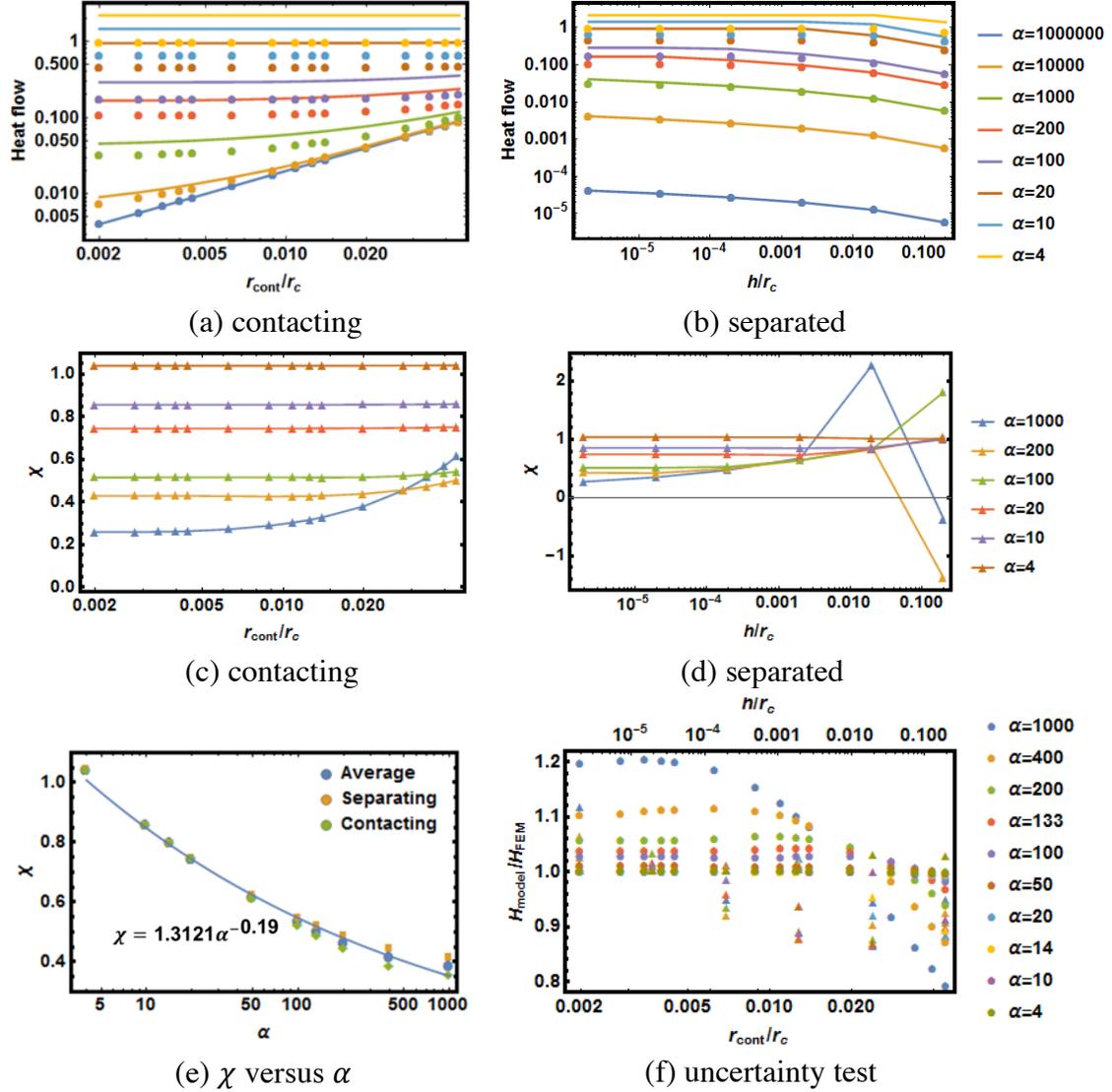

(a) contacting
(b) separated
(c) contacting
(d) separated
(e) $\chi$ versus $\alpha$
(f) uncertainty test

Figure 3 (a) and (b) compare the heat flow between the finite element simulation (filled circles) and the original Batchelor & O'Brien model (solid lines), coloured according to $\alpha$. (c) and (d) plot $\chi$ for each $\alpha$ and $r_{\text{cont}}/r_c$ pair as well as $\alpha$ and $h/r_c$ pair with joining lines, respectively. (e) gives the correlation between $\chi$ and $\alpha$ and (f) compares the computed heat flow between the modified model and the simulated result.

The heat flow calculated by the modified BOB model with the proposed correlation presents reasonable agreement with the finite element simulation results, but large $\alpha$ reduces its accuracy. The $\chi$ calculated from this function for $\alpha = 120$ is 0.528, close to 0.5 as used in the literature. While $\alpha = 10$ case gives a larger discrepancy, 0.847 versus 0.71, probably due to the polydispersity in that research [26]. Nevertheless, this function is useful for predictive purposes and the modified (BOB) model thus has good applicability, especially for conditions of $\alpha <$ 1000. For $\alpha > 10000$, the original BOB model is preferable.

2.3 **Inclusion of the Smoluschowski effect**

The Smoluschowski effect relates to a reduced thermal conductivity through gases over scales smaller than the mean free path of gas molecules, as the result of boundary effects at a solid-gas interfaces. This effect becomes significant in granular materials where a gaseous phase sits in gaps between surfaces separated by a short distance, [46]. This effect brings about a dependence of thermal conductivity on gas pressure and grain size [17, 26], and contributes to the deviation of $\chi$ as discussed in the previous section. In order to improve the accuracy of the modified model, the Smoluschowski effect is implemented using the method described below [26].

The influence of the Smoluschowski effect on the thermal conductivity of gases is described by

$$k'_g = \frac{k_g}{1+2\,\gamma\,\text{Kn}}, \qquad \text{2-6}$$

where $\gamma$ quantifies the efficiency of energy transfer between gas molecules and the solid surface, and is given in terms of the molar mass ratio of the solid over gas with $m_r = m_s/m_g$,

$$\gamma = \frac{2\,(1+m_r)^2 - 2.4\,m_r}{2.4\,m_r}, \qquad \text{2-7}$$

and the Knudsen number Kn is determined by the feature size $L_f$ of given space and the mean free path (MFP) of gas molecules as,

$$\text{Kn} = \frac{\text{MFP}}{L} = \frac{k_B\,T}{\sqrt{2}\,\pi\,d_g^2\,P_g\,L}, \qquad \text{2-8}$$

where the gas phase is characterised by pressure $P_g$, thermodynamic temperature $T$, the kinetic molecular diameter $d_g$ and the Boltzmann constant $k_B$.

The quantities required to determine $k'_g$ are gas related except for the feature size of space $L_f$. The geometry of contact units is used to obtain $L_f$ as following. In the contacting scenario,

$$L_{\mathrm{f}} = r_i \left( \frac{\sin^{-1}\left(\frac{r_{\min}}{r_i}\right) - \sin^{-1}\left(\frac{r_{\mathrm{cont}}}{r_i}\right) - \frac{r_{\min}}{r_i} + \frac{r_{\mathrm{cont}}}{r_i}}{\sin^{-1}\left(\frac{r_{\min}}{r_i}\right) - \sin^{-1}\left(\frac{r_{\mathrm{cont}}}{r_i}\right)} \right) + r_j \left( \frac{\sin^{-1}\left(\frac{r_{\min}}{r_j}\right) - \sin^{-1}\left(\frac{r_{\mathrm{cont}}}{r_j}\right) - \frac{r_{\min}}{r_j} + \frac{r_{\mathrm{cont}}}{r_j}}{\sin^{-1}\left(\frac{r_{\min}}{r_j}\right) - \sin^{-1}\left(\frac{r_{\mathrm{cont}}}{r_j}\right)} \right).$$
2-9

and for separated particles,

$$L_{\mathrm{f}} = h + r_i \left( \frac{\sin^{-1}\left(\frac{r_{\min}}{r_i}\right) - \frac{r_{\min}}{r_i}}{\sin^{-1}\left(\frac{r_{\min}}{r_i}\right)} \right) + r_j \left( \frac{\sin^{-1}\left(\frac{r_{\min}}{r_j}\right) - \frac{r_{\min}}{r_j}}{\sin^{-1}\left(\frac{r_{\min}}{r_j}\right)} \right).$$
2-10

$r_{\min}$ is set as the minimum between $r_i$ and $r_j$ in one unit for a more general purpose to include poly-dispersed granular media.

For the contacting grain-wall pairs,

$$L_{\mathrm{f}} = r_i \left( \frac{\frac{\pi}{2} - \sin^{-1}\left(\frac{r_{\mathrm{cont}}}{r_i}\right) - 1 + \frac{r_{\mathrm{cont}}}{r_i}}{\frac{\pi}{2} - \sin^{-1}\left(\frac{r_{\mathrm{cont}}}{r_i}\right)} \right).$$
2-11

and the separated grain-wall pairs,

$$L_{\mathrm{f}} = r_i \left( \frac{\frac{\pi}{2} - 1}{\frac{\pi}{2}} \right).$$
2-12

Eventually, $\alpha = k'_{\mathrm{g}}/k_{\mathrm{s}}$ is derived and used for the modified BOB model.

## 2.4 Simulation protocol

The grain-scale heat transfer model is coded in C++ within LIGGGHTS framework and is provided in supporting information. For each simulation, a granular medium is generated by the free-falling method in a given volume with specified boundary conditions. Following particle settling under gravitational force onto the flat bottom surface of the container, a flat upper surface is added and is used to apply mechanical compression. The upper surface is set as a heat source while the bottom wall is treated as a heat sink with a constant temperature gap $\Delta T_{\mathrm{gap}}$ of 1K between these two boundaries. When a target loading is achieved, the granular medium is frozen, the walls are locked, and adiabatic boundary conditions are imposed in lateral directions. The granular medium is then heated until the simulation reaches steady states where the fractional difference between inflow heat (top) and outflow heat (bottom) is smaller than $10^{-6}$. Finally, the effective thermal conductivity $k_{\text{eff}}$ is expressed as

$$k_{\text{eff}} = \frac{H_{\text{inflow}} L_{\text{axis}}}{A_{\text{cross-section}} \times \Delta T_{\text{grap}}}. \qquad 2\text{-}13$$

where $H_{\text{inflow}}$ indicates the inflow heat, and $L_{\text{axis}}$ and $A_{\text{cross-section}}$ are respectively the height and the cross-sectional area of the granular medium. Alternatively, $k_{\text{eff}}$ can be calculated also as

$$k_{\text{eff}} = \frac{H_{\text{inflow}}}{A_{\text{cross-section}}} \times \frac{1}{\Delta T_{\text{gradient}}}. \qquad 2\text{-}14$$

with $\Delta T_{\text{gradient}}$ representing the temperature gradient inside the granular medium. In general, by binning the grains according to their height, a linear function can be fitted based on the mean temperature of each bin against the representative height of each bin. Thus, $\Delta T_{\text{gradient}}$ can be calculated as the slope of that linear function. Compared with the former expression, using $\Delta T_{\text{gradient}}$ can exclude the wall-grain contacts and gives the more inherent $k_{\text{eff}}$ of the granular medium. However, the former one is considered more appropriate for experimental validation as practical measurement always includes wall-grain contacts.

## 3 Experimental validation and discussion

The HotDisk system built using a transient plane source technique [47, 48] is employed to experimentally measure the effective thermal conductivity of granular media. Two kinds of grains, 2 mm glass ($k_s = 1.2$ W/mK) beads and 2 mm 304 stainless steel ($k_s = 16.2$ W/mK) beads, are used separately to form solid-gas media in air under ambient conditions of atmospheric pressure and room temperature ($k_g = 0.026$ W/mK). The Smoluschowski effect is described by the correlations shown in Table 1. A load-bearing sample container is designed to hold these granular media. In this way, the mechanical influence on effective thermal conductivity of granular media can be investigated.

Measurement results, shown in Figure 4, reveal that the thermal transport of the two granular systems responds differently to mechanical loading. The granular medium consisting of glass beads is insensitive to mechanical loading while the $k_{\text{eff}}$ of the packed steel grains increases gradually with load. This difference can be understood in light of the fact that steel has a thermal conductivity more than ten times higher than that of glass. Larger load creates larger intergranular contact surfaces, which results in greater heat conduction pathways, with these having greater significance in the more thermally conductive steel. Discrete element simulation results are quantitatively similar, demonstrating that the applicability of the present heat transfer model.

Table 1 Gas properties used in this work

| Gas type | Properties | |
|---|---|---|
| Helium (He) | $k_g$ / (W/mK): | $3.366 \times 10^{-3} T^{0.668}$ ($T$ in K). |
| | $d_g$ / (m): | $2.15 \times 10^{-10}$. |
| | $m_g$ / (g/mol): | 4. |
| Air | $k_g$ / (W/mK): | $-10^{-11} T^3 - 4 \times 10^8 T^2 + 8 \times 10^5 T + 0.0241$ ($T$ in °C). |
| | $d_g$ / (m): | $3.66 \times 10^{-10}$. |
| | $m_g$ / (g/mol): | 28.96. |

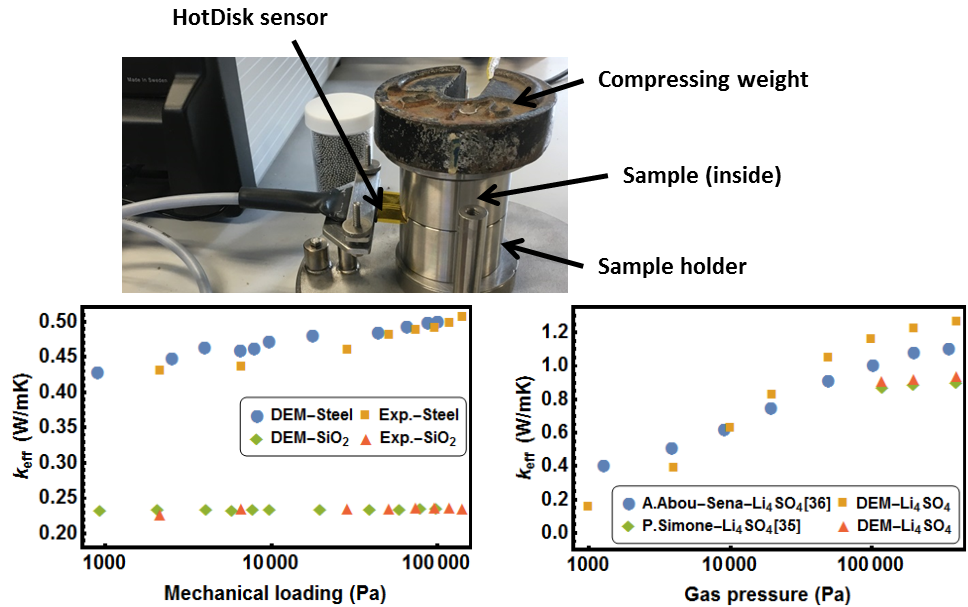

Figure 4 Top – Experimental configuration. Bottom – Comparison between simulation results and the experimental measurement in this work (left) and previously published studies (right).

In addition, the role of the Smoluschowski effect and poly-dispersed particle size distributions are also examined with reference to published experimental results. A poly-dispersed granular medium of lithium orthosilicate Li$_4$SiO$_4$ [49] ($k_s$ = 2.57 W/mK at 20 °C and $k_s$ = 2.15 W/mK at 500 °C) with helium filling ($k_g$ = 0.15 W/mK at 20 °C and $k_g$ = 0.26 W/mK at 500 °C) is fed to the discrete element simulation. $k_{\text{eff}}$ of the simulated media shows results closely matching experimental data including gas pressure dependence. It should be pointed out that this heat transfer simulation slightly overestimates $k_{\text{eff}}$ when compared with experimental data [50] despite neglecting thermal radiation.

## 4 Effective thermal conductivity vs. packing structure

The role of material properties and grain-grain interactions in overall thermal transport have been examined through the implementation of the discrete element simulation reported here, and have also been widely reported in the literature. Therefore, the following sections focus on establishing relationships between packing order and effective thermal conductivity.

### 4.1 Packing structure transition by vibration

It has been demonstrated that mechanical vibration can effectively change the packing structure of granular media [51, 52]. For granular media settling freely under gravity, random disordered packing structures form with a few clusters of hexagonal-close-packed (HCP) and face-centred-cubic (FCC) packings. By vibrational agitation of granular media disorder-to-order transitions are provoked. Grains gradually rearrange themselves into structured clusters to achieve dynamic efficiency [41, 53]. Not only can the packing fraction be varied, but also the order level of the packing structure significantly increases during vibration. The vibration protocol used in this study is as follows: Granular media with a constant number of grains are generated by the free-falling method within finite cylinders of varying diameters under gravitational force using LIGGGHTS. These packings are then subjected to continuous mechanical vibration by sine wave agitation of the bottom surface. As a result, grains are excited and relocate into new positions, forming structured clusters that grow and finally reach quasi-equilibrium states [53]. Three types of granular media are simulated and their geometrical dimensions are summarised in Table 2. For each geometry, five samples are extracted at different vibration durations for subsequent heat transfer simulation.

Table 2 Geometry parameters of granular media

| | |
|---|---|
| Diameter of grain | 2.3 (mm) |
| Number of grains | 5000 |
| Cylinder Diameter/Height | 30/75, 40/60, 50/25 (mm/mm) denoted as D30, D40, D50 |

To quantify this disorder-to-order transition, the structural index $S_6$ is utilised to measure the rotational symmetry of the neighbour configuration of individual grains. In such measurement, the neighbour configuration of each grain is first represented by a set of vectors pointing from a given grains to its neighbours, i.e., 12 nearest grains in current study, and then the cosine similarity between the vector set of a particular grain and those vector sets of its neighbours is used to evaluate the resemblance between the neighbour configurations of these grains. Thus, the order level of grain-scale packing structure is positively described by the aforementioned resemblance, with values ranging from 0 to 12 [39, 53-55]. Voronoi tessellation [56] is employed to obtain the effective volume $V_V$ and packing fraction $\varphi_V$ of individual grains. In this way, the mean values of $S_6$ and $\varphi_V$, excluding surface grains, can be recognised as the global characteristics of a granular medium. As shown in Figure 5, the mean values of both of $S_6$ and $\varphi_V$ monotonically increase with vibration, which implies a cooperative process between densification and ordering in granular media. The reason for the sudden increase of $\varphi_V$ in D50 can be explained by the geometrical effect in the ordering process [53]. The large cylinder diameter promotes the growth of perfect HCP and FCC clusters, while the small one causes distortion in the ordered clusters with imperfection.

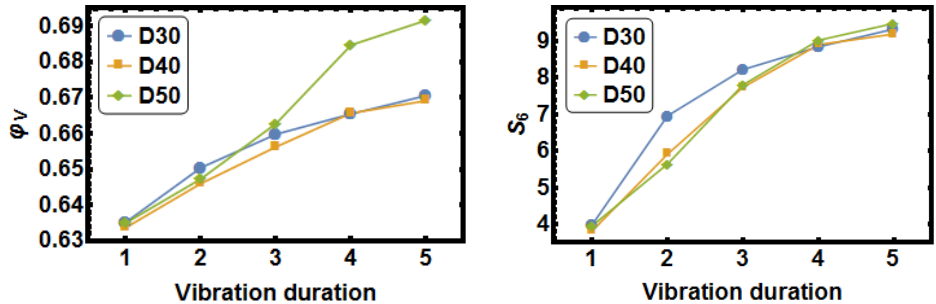

Figure 5 $\varphi_V$ (– left) and $S_6$ (– right) increase as the duration of vibration is extended, noticing that the horizontal axis is arbitrary.

By considering individual grain positions, the spatial resolution of $S_6$ can be further revealed. Due to the symmetry of the cylindrical containers, two-dimensional (2D) radial sections are studied and sub-domains are created by mapping a 2D grid with grain diameter resolution onto these sections. By sweeping the cylindrical spaces with the 2D sections, grains are mapped into the relevant sub-domains according to their radial distance $x_{radial}$ and height $z$, and then the mean $S_6$ of each sub-domain can be obtained to yield intensity maps of $S_6$. Spatial patterns of these maps are found to be nearly axisymmetric. A typical evolution of $S_6$ distribution is shown in Figure 6 presenting a spatially inhomogeneous order development.

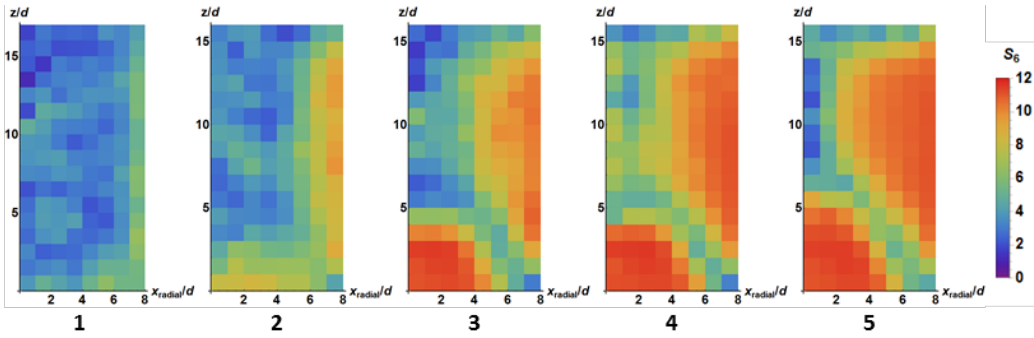

Figure 6 Evolution of $S_6$ spatial patterns in D40 with vibration duration (from left to right).

### 4.2 Enhanced effective thermal conductivity by vibration

To examine the role of packing structure in thermal conductivity we consider a medium consisting of glass ($SiO_2$) grains filled with a gas phase of helium (He). By varying the helium pressure in the simulation, $\alpha$ can be varied over a large range for a given solid phase. According to Figure 7 (a), the Smoluschowski effect strongly influences the thermal conductivity of He in the pressure range from 1Pa to $10^5$ Pa and brings about a change in $\alpha$ roughly from $10^4$ to 10, respectively. Due to the correlation between $\chi$ and $\alpha$, helium pressures of 10 Pa, $10^2$ Pa, $10^3$ Pa, $10^4$ Pa and $10^5$ Pa are used. To avoid floating grains, granular media are initially compressed by a mechanical loading of 1 MPa.

By using the fitted $\Delta T_{gradient}$ in Eqn. (2-14), the boundary effect is subtracted and a more inherent value of $k_{eff}$ can be obtained. Firstly, correlations between $k_{eff}$ and the global $\varphi_V$ as well as between $k_{eff}$ and the global $S_6$ are studied. In Figure 7, (b) and (c) show plots of the minimum $\alpha$ ($\approx 1000$) while (d) and (e) are extracted from the maximum $\alpha$ ($< 10$). A collective correlation between $k_{eff}$ and the global $\varphi_V$ of different samples can be observed for large $\alpha$, but the correlations split at small $\alpha$ values. In contrast, correlations between $k_{eff}$ and the global $S_6$ tend to converge for small $\alpha$ but diverge for large $\alpha$. This difference indicates that the corresponding influences of densification and ordering are gradually weakened and strengthened, respectively, as $\alpha$ decreases. Besides, the unusual drop and jump in (b) and (c) should be clarified. The breakdown of the monotonic increase of $k_{eff}$ against $\varphi_V$ is considered as the result of insufficient contact between the heat source wall and boundary grains in the simulation, leading to poorer heat transfer into the media. However, this influence become insignificant when $\alpha$ is small.

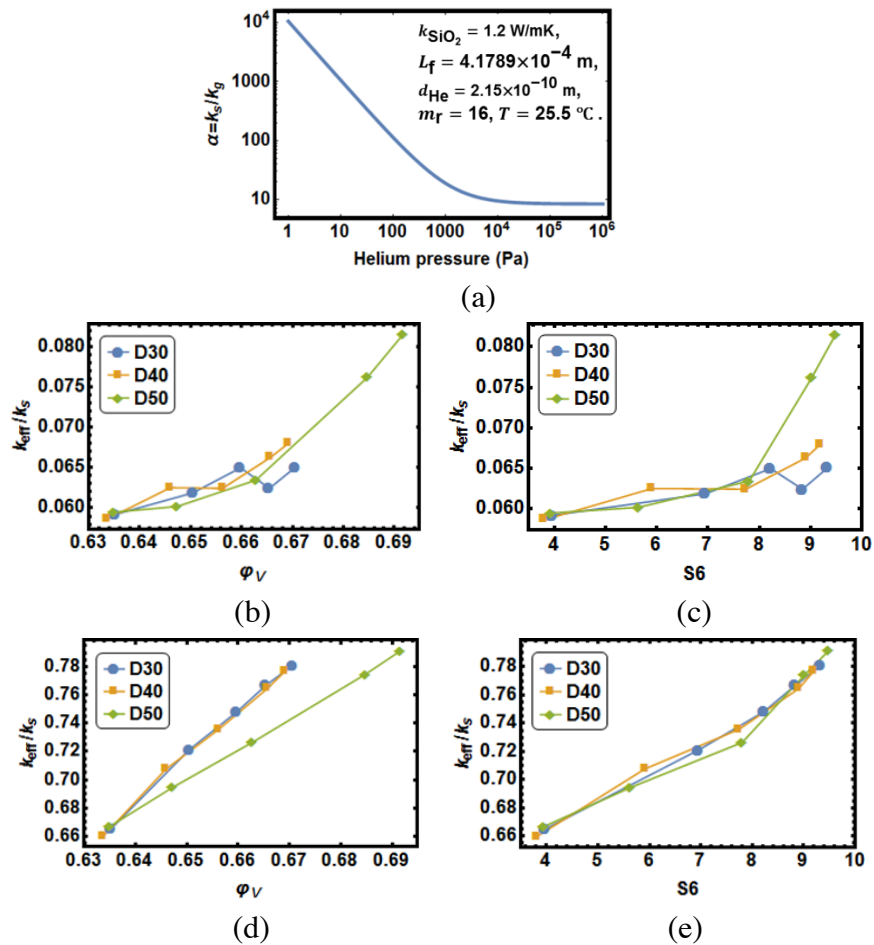

Figure 7 (a) The variation of $\alpha$ with helium gas pressure in SiO$_2$-He granular media consisting of 2.3-mm-diameter SiO$_2$ grains at room temperature. (b) and (c) show the correlations of ($k_{eff}$, $\varphi_V$) and ($k_{eff}$, $S_6$) in the minimum $\alpha$ ($\approx 1000$), respectively. (d) and (e) show similar correlations of the maximum $\alpha$ ($< 10$).

This global observation indicates that neither packing fraction nor structure order can determine $k_{\text{eff}}$ alone. The collective processes of densification and ordering are the reason for the enhancement of $k_{\text{eff}}$ introduced by vibration. But, by decreasing $\alpha$, the prominence in such processes shifts from packing fraction to structure order.

### 4.3 Grain-scale structure and heat transfer

It has been shown that $k_{\text{eff}}$ of granular media is positively correlated to both $\varphi_V$ and $S_6$ at the macroscale, but mechanisms at finer scales are insufficiently understood. As demonstrated in Figure 6, the structural transition brings about substantial and inhomogeneous grain-scale ordering. Whether this improvement enhances heat transfer in granular media and if so, how this enhancement works remain elusive. In order to address this interplay between grain-scale structure and thermal transport, analysis of heat transfer at the grain-scale is performed.

Analytically, a heat flux density vector $\boldsymbol{H}_{\text{density}}$ across the surface of an individual grain in a granular medium can be presented as [13],

$$\boldsymbol{H}_{\text{density}} = n \sum_i \boldsymbol{x}_i H_i. \qquad 4\text{-}1$$

where $H_i$ is the heat flow passing through each contact, $\boldsymbol{x}_i$ is the vector pointing from the centroid of the particular grain toward the corresponding contact point and $n$ is the grain number density, which can also be calculated at the grain-scale as the reciprocal of the Voronoi volume [56], $1/V_V$ of the corresponding grain. Further, it has also been proved that the temperature gradient vector of each grain is equal to the global temperature gradient vector $\Delta \boldsymbol{T}_{\text{gradient}}$ of the entire granular medium [13]. Based on this consideration, an expression is validated for each grain,

$$\boldsymbol{H}_{\text{density}} = \boldsymbol{k}^* \Delta \boldsymbol{T}_{\text{gradient}} = n \sum_i \boldsymbol{x}_i H_i. \qquad 4\text{-}2$$

with $\boldsymbol{k}^*$ being the second rank thermal conductivity tensor for each grain. Therefore, $\boldsymbol{k}^*$ of individual grains can be derived if the corresponding neighbour configuration and the heat flows through these neighbouring grains are known. Such information can be easily extracted from the discrete element simulation, which facilitates this grain-scale heat transfer analysis. Besides, the sum of heat flows of each grain is 0 at steady-state, $\sum_i H_i = 0$, so by separately summing inflow or outflow heat for each grain, the scalar quantity of total heat flow $H_{\text{flow}}$ through one grain can also be calculated. In such measurements, two aspects of heat transfer at the grain-scale can be investigated, the magnitude of the heat flow and the directional thermal conductivity. Thus, the fitted $\Delta T_{\text{gradient}}$ in Eqn. (2-14) is used and resulted in $\Delta \boldsymbol{T}_{\text{gradient}} = \{0, 0, \Delta T_{\text{gradient}}\}$ due to the imposed temperature boundary condition. Because of the uni-directional nature of the temperature gradient, only the $z$ component of thermal conductivity tensor $\boldsymbol{k}^*$ is needed, giving $\{k_{zx}, k_{zy}, k_{zz}\}$. The principle component $k_{zz}$ is of most interest, and the anisotropy $\theta_z = \text{ArcCos}(k_{zz}/|\boldsymbol{k}^*|_z)$ of this $z$ component of $\boldsymbol{k}^*$ is examined here.

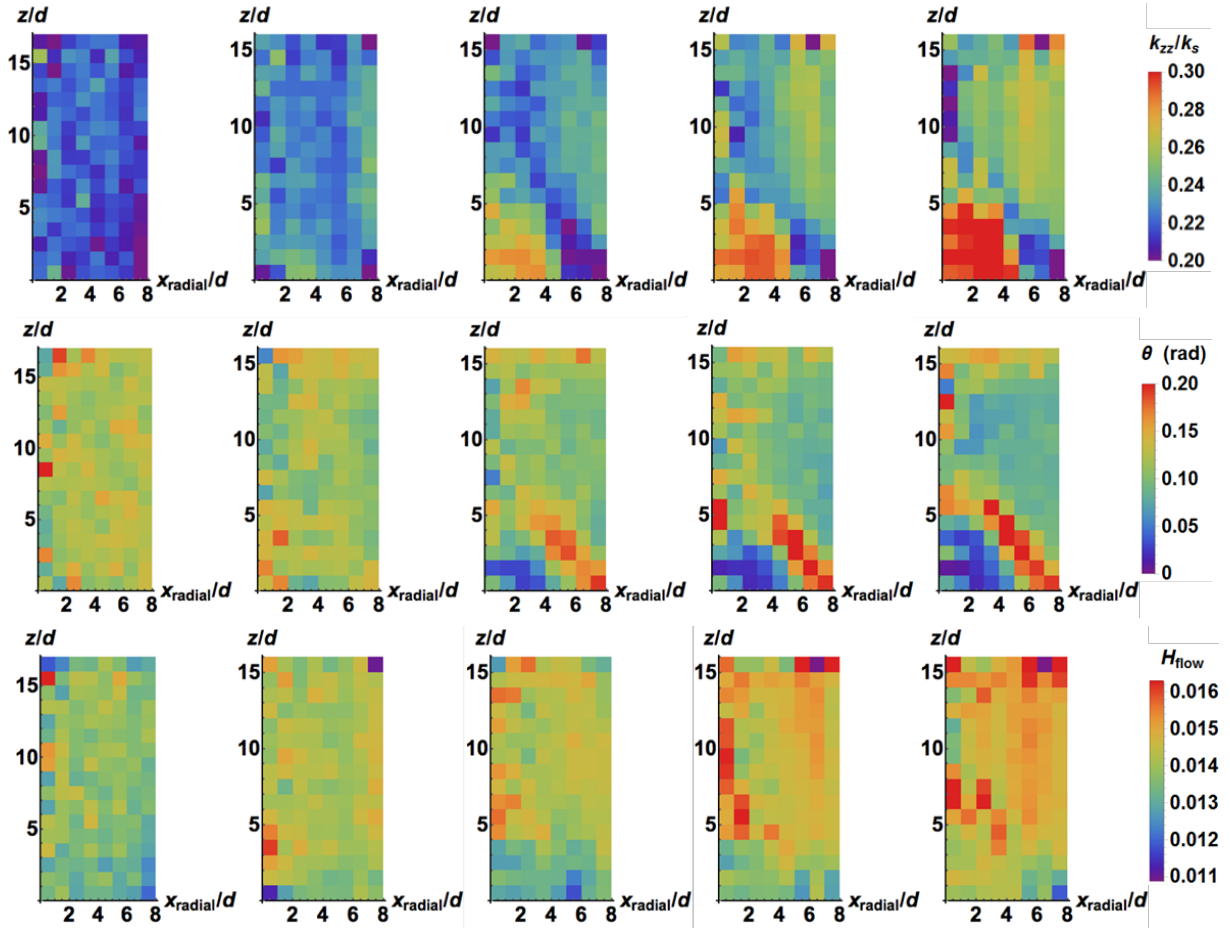

Figure 8 Spatial evolution of $k_{zz}/k_s$ (top row), $\theta$ (middle) and $H_{\text{flow}}$ (bottom) of the D40 granular media.

A similar analysis to that presented in Figure 6 is applied to examine the spatial distributions of the thermal conductivity tensor $k_{zz}$, its anisotropy $\theta_z$ and the grain level heat flow ($H_{\text{flow}}$). The spatial evolution of the ratio $k_{zz}/k_s$ and $\theta_z$ for a D40 sample with $\alpha \approx 10$ is shown sequentially in the top and middle rows of Figure 8, respectively. The spatial distribution of $k_{zz}/k_s$ correlates positively to that of $S_6$ in Figure 6, while the patterns of $\theta_z$ exhibit an inverse trend. In contrast, the spatial distribution of non-dimensional $H_{\text{flow}}$ (bottom row in Figure 8) does not show any observable trend that can be correlated to structural patterns, and exhibits only a globally increasing trend along with the overall $S_6$, as vibration proceeds. This suggests that $k_{zz}$ and $\theta_z$ are closely related to the grain-scale structure but $H_{\text{flow}}$ is less influenced.

Comparing the evolution of the spatial distributions of $k_{zz}/k_s$ and $\theta_z$ with the evolution of $S_6$, it can be seen that, at the grain-scale, ordering increases thermal conductivity and decreases anisotropy, resulting in similar inhomogeneous patterns. In fact, due to the cooperativity of the densification and ordering, the spatial patterns of $\varphi_V$ are nearly identical to those of $S_6$. Thus, it is hard to differentiate the influences of these two processes, but ordering is considered to be precedence of densification in granular media due to the rigidity of grains. However, it is clear that the decrease of anisotropy, $\theta_z$, can be related to a greater symmetry of the neighbourhood configurations of individual grains, which is specifically measured by $S_6$. Further, the bottom ordered region in the patterns of Figure 6 exhibits a higher $k_{zz}/k_{\text{eff}}$ and lower $\theta_z$ in Figure 8 than the corresponding side ordered region. This contrast can be attributed to the different completeness of ordering in these two regions. As mentioned before, due to the curved cylinder wall, distortion occurs in the side region [53], making the ordered packing structures deviate from perfect FCC and HCP and become looser as well as less symmetric. In summary, two grain-scale mechanisms to enhance the macroscopic $k_{\text{eff}}$ are identified regarding to the ordered packing structure.

By transforming the quantitative results from the mapping operations of Figure 5 and Figure 8 into the coordinate plots shown in Figure 9, parameter correlations are further investigated. The Pearson correlation coefficient is employed to quantitatively evaluate the correlations of ($S_6$, $k_{zz}$) and ($S_6$, $\theta_z$) in the left and middle of Figure 9, respectively. It confirms the positive correlation between $S_6$ and $k_{zz}$ together with the negative correlation between $S_6$ and $\theta_z$.

However, such correlations become stronger as the diameter of the granular media increases. The reason for this augmentation can be understood by considering the aforementioned differences in the structured regions; because the packing structures formed in D30 are dominated by the distorted HCP structures [53] while D50 consists of perfectly ordered structures. In addition, the correlation between $\varphi_V$ and $k_{zz}$ is found to give the highest Pearson correlation coefficient, although this is unsurprising as $\varphi_V$ is used in the calculation of $k_{zz}$ in Eqn. (4-2).

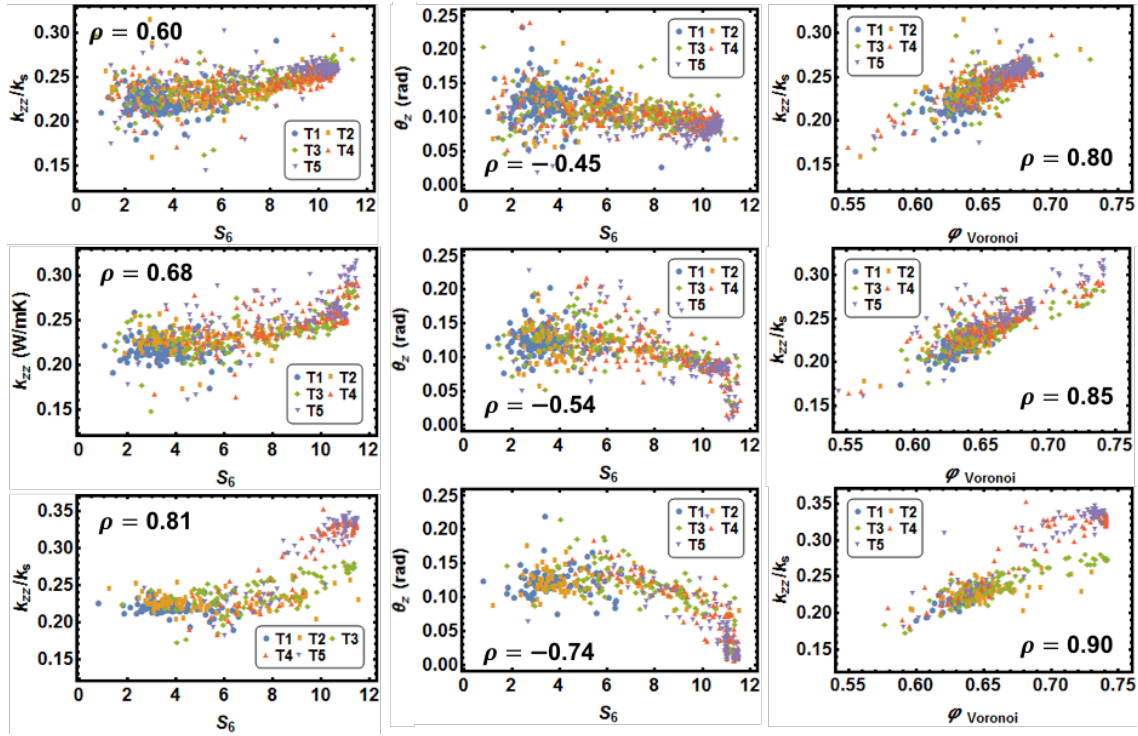

Figure 9 Left – ($S_6$, $k_{zz}$), middle – ($S_6$, $\theta_z$) and right – ($\varphi_V$, $k_{zz}$) plots of D30, D40 and D50 (from top to bottom respectively), created to according to the mapping operation, in the scenario of $\alpha = 100$.

By employing a similar method, the influence of $\alpha$ observed in Figure 7 is further examined. The ($S_6$, $k_{zz}$) and ($S_6$, $\theta_z$) coordinates are plotted in Figure 10 for well-structured granular media following the final vibration time. Lower values of $\alpha$ enhance the various correlations found here. This arises as the reduction of $\alpha$ enlarges the effective volume of gas phase participating in heat transfer at contacts, achieving a more homogeneous heat flow profile on grain surfaces, which compensate for the structural distortion that arises from the cylindrical boundary. This effect can be implied from the contrast, especially the significant decreases of $\theta_z$ between (b) and

(d) in D30 and D40 samples, as the $\alpha$ drops from 1000 to 10. Due to the relatively poor order of the side region in D50 samples regions [53], the compensating effect by the decrease of $\alpha$ becomes less effective. This discussion indicates an appropriate interpretation of the order of packing structure is necessary in the prediction of $k_{\text{eff}}$ of granular media with small $\alpha$.

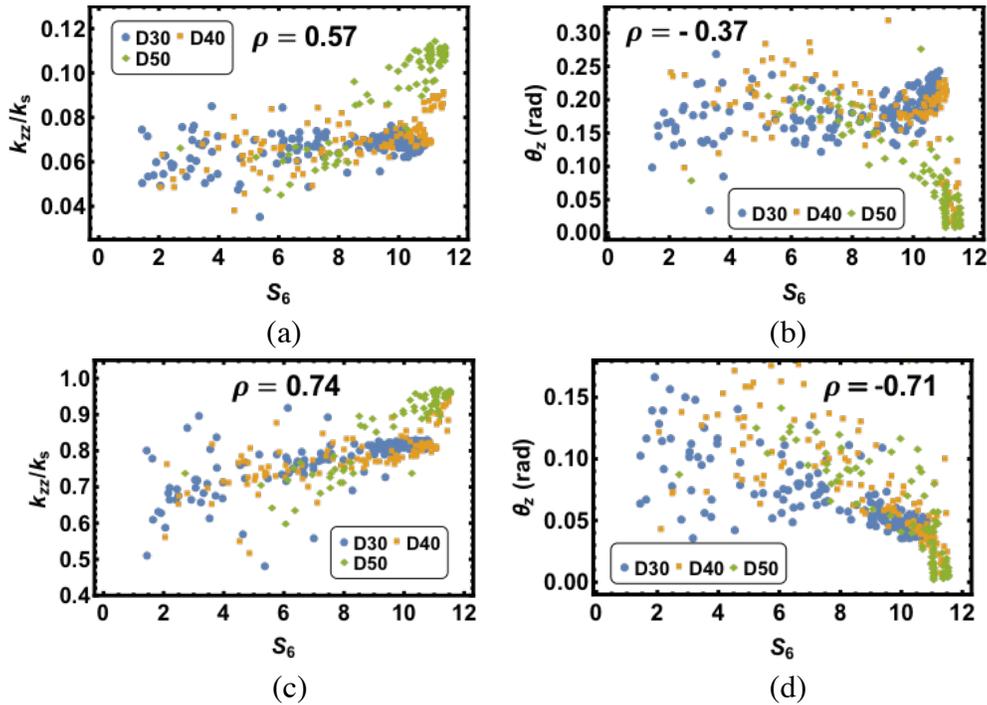

Figure 10 ($S_6$, $k_{zz}$) and ($S_6$, $\theta$) plots of well-ordered packing structure belonging to three types of granular media in different scenarios. (a) and (b) – $\alpha \approx 1000$, (c) and (d) – $\alpha \approx 10$.

Finally, it is necessary to point out that it is challenging to untangle the effects of packing ordering and densification in granular media. Nonetheless, an increase in local ordering can be clearly attributed to two results. One is the enhancement of grain-scale thermal conductivity along with the increase of packing fraction. The second is the reduction of anisotropy, improving the alignment of grain thermal conductivity with the principle heat transfer direction. Further, the effects of ordering would be more pronounced in systems where not all grains are in contact e.g. cemented granular media and grain-filler composites. To check this point, the packing structures of those granular media are reused to form artificial media, which can possibly be considered as grain-filled composites. In these composites, the grains are fixed at the identical positions with the order level maintained (disorder – initial state and well-order – longest vibrated state, corresponding to Figure 5), but the radii of the grains are varied to adjust the packing fraction alone. As shown in Figure 11, two scenarios of $\alpha = 1000$ (left) and $\alpha = 10$ (right) are presented, where the Smoluschowski effect is turned off to avoid the size dependency due to the change of grain radius.

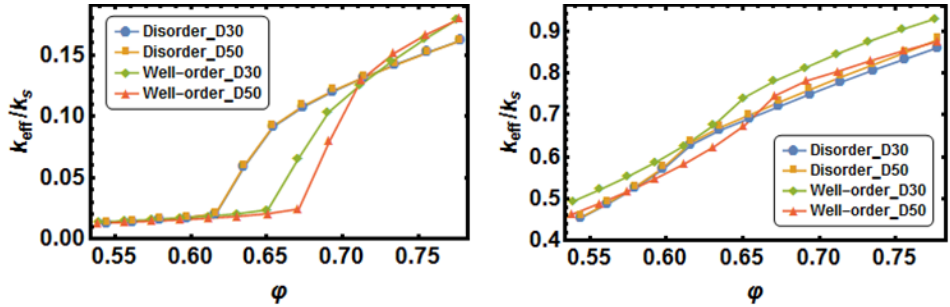

Figure 11 The variation of $k_{\text{eff}}$ against the change of packing fraction in artificial media with disorder and well-order packing structures in $\alpha = 1000$ (left) and $\alpha = 10$ (right) scenarios.

Firstly, in spite of the dissimilar varying tendencies of $k_{\text{eff}}$ against $\varphi$ in Figure 11, a comparable shape of S can be identified. Although the S shape is flattened in the condition of $\alpha = 10$, a transition step is still observed. This invariant transition reflects the formation and loss of contacts between grains in corresponding to the change of packing fraction, i.e., the grain radius. which is the reason for the sharp transition in the condition of $\alpha = 1000$. Regarding to the effect of the structural order, it works differently in according to $\alpha$. In the large $\alpha$ condition, the ordered structure has relative uniform separation distance between grains, making the plunge of $k_{\text{eff}}$ become sharper than the disorder structure. But higher coordination number of the ordered structure improves the possibility of contact formation, resulting in higher $k_{\text{eff}}$ with the packing fraction increasing. The mergence as the reduction of packing fraction in this scenario can be considered as the loose dispersion limit of the Maxwell model [8] where the structural order is least influential. However, in the condition of small $\alpha$, the preference between packing order and packing fraction further depends on the detailed packing structure that is influenced by the geometry. Due to the presence of the compensation by small $\alpha$, the slender medium (D30) of ordered structure generally exhibits higher $k_{\text{eff}}$ in the present packing fraction range. On the contrary, the flat medium (D50) presents similar change of the preference to the one of $\alpha = 1000$. Such difference has been attributed to the poorly ordered side region, which can also be recognised as the segregation of disordered region. This segregation introduces disorder of larger dimension beyond the grain-scale, so out of the scope of current study.

## 5   Conclusion

In this work, a bi-phasic heat transfer model based on the Batchelor & O'Brien solutions was successfully implemented in the open source software LIGGGHTS. To overcome the inaccurate prediction of the original model when the assumption $k_s/k_g \gg 1$ is unsatisfied, a compensation method with only one variable $\chi$ is applied. With the help of finite element analysis of the heat transfer process of a series of contact geometries and $k_s/k_g$ ratios, an empirical fitting function is proposed to determine $\chi$ according to $k_s/k_g$ for individual contacts. For further accuracy, the Smoluschowski effect is included in the modified model. Finally, this heat transfer simulation framework is validated by experimental measurements employing a transient plane source technique as well as comparison with literature data. The simulated results show good agreement with mono-dispersed and poly-dispersed granular media, while deviation becomes larger when grain size distributions are unknown. As thermal radiation and convection have been ignored, the applicability of this framework is optimal for granular media with stagnant gas phases at room temperature.

Granular media comprising glass and helium with different packing structures are applied in this heat simulation framework to study how the order characteristics of packing structure affect the effective thermal conductivity of these media. By investigating thermal properties and order characteristics at the grain-scale, it was demonstrated that the structural disorder-to-order transition enhances the overall thermal conductivity $k_{\text{eff}}$. Locally, the increase of thermal conductivity in the principal direction, $k_{zz}$, and the decrease of anisotropy $\theta_z$, are further identified as outcomes of the structural transitions, which result from grain-scale ordering processes. The heat flow through individual grains, as indicated by $H_{\text{flow}}$, is found to correlate strongly with the global structural variation but does not exhibit a clear relationship with local structure. The two grain-scale effects are further influenced by the $k_s/k_g$ ratio. The findings here provide a potential method to manage heat transfer in granular media through manipulation of packing structure and provide a conceptual framework for the broader analysis of transport phenomena in granular media and heterogeneous media in general.